\documentclass{caosp306}
\usepackage{graphicx}
\usepackage{natbib}
\usepackage{amsmath,amssymb,enumerate}
\articleNo{123}
\pubyear{2005}
\volume{35}
\volnumber{3}
\firstpage{1}
\received{\date{}}
\accepted{\date{}}
\def\BibTeX{{\rm B\kern-.05em{\sc i\kern-.025em b}\kern-.08em
             T\kern-.1667em\lower.7ex\hbox{E}\kern-.125emX}}

\begin{document}
\renewcommand{\makeheadbox}{}

\htitle{Possible observational signatures of SMBHBs in their Fe K$\alpha$ line 
profiles}
\hauthor{P.\,Jovanovi\'{c} {\it et al.}}

\title{Possible observational signatures of supermassive black hole binaries in their Fe K$\alpha$ line profiles}

\author{
        P. Jovanovi\'{c} \inst{1,*} 
      \and 
        V. Borka Jovanovi\'{c} \inst{2}   
      \and 
        D. Borka \inst{2}
      \and
        L. \v{C}. Popovi\'{c} \inst{1}
       }

\institute{
           Astronomical Observatory, Volgina 7, P.O. Box 74, 11060 Belgrade, Serbia $^*$\email{pjovanovic@aob.rs}
         \and 
           Atomic Physics Laboratory (040), Vin\v{c}a Institute of Nuclear Sciences, University of Belgrade, P.O. Box 522, 11001 Belgrade, Serbia
          }
\date{March 8, 2003}

\maketitle

\begin{abstract}
Here we study the potential observational signatures of supermassive black 
hole binaries (SMBHBs) in the Fe K$\alpha$ line profiles emitted from the 
accretion disks around their components. We simulated the Fe K$\alpha$ line 
emission from the relativistic accretion disks using ray tracing method in Kerr 
metric. The obtained profiles from the SMBHBs are then compared with those in 
the case of the single supermassive black holes (SMBHs). We considered two 
models of the SMBHBs: a model when the secondary SMBH is embedded in the 
accretion disk around the primary, causing an empty gap in the disk, and a 
model with clearly separated components, where the accretion disks around both 
primary and secondary give a significant contribution to the composite Fe 
K$\alpha$ line emission of a such SMBHB. The obtained results showed that both 
models of SMBHBs can leave imprints in the form of ripples in the cores of the 
emitted Fe K$\alpha$ line profiles, which may look like an absorption 
component in the line profile. However, in the case of the composite line 
profiles emitted from two accretion disks, these ripples could have much higher 
amplitudes and strongly depend on orbital phase of the system, while for those 
emitted from a disk with an empty gap, the corresponding ripples mostly have 
lower amplitudes and do not vary significantly with orbital phase. The present 
day X-ray telescopes are not able to detect such signatures in the observed 
X-ray spectra of SMBHBs. However this will be possible with the next generation 
of X-ray observatories, which will also enable application of such effects as a 
tool for studying the properties of these objects.
\end{abstract}

\section{Introduction}

Nowadays, it is widely accepted that binary systems of SMBHs originate in 
galactic mergers \citep{beg80}, and that their coalescences are the powerful 
emitters of low-frequency (nHz) gravitational waves (GWs) which are currently 
probed by pulsar timing arrays (PTA) \citep{ses18}, and which will be 
main targets for future space-based interferometers. Searches for (active) 
SMBHBs are currently ongoing, resulting with more than 100 candidates 
\citep[see e.g.][and references therein]{ses18}, and were significantly 
intensified after the first observation of merging stellar mass black hole 
binary performed by \citet{ligo16}.

One of the most powerful methods for SMBHB searches is the spectroscopy (in 
different spectral bands). As two SMBHs in a galactic merger become 
gravitationally bound and start to orbit around their center of mass, the 
emission lines from SMBH components start to shift due to their radial 
velocities \citep[see e.g.][for examples in optical band]{pop12,bon12,li16}. 
In such a case, a strong X-ray emission in the broad Fe K$\alpha$ line at 6.4 
keV could arise from accretion disks around both SMBHs, and could be therefore 
affected by the Doppler shifts due to the orbital motion of the binary 
\citep{yulu01,jova14}, since the radial velocities of its components could
reach $\approx1.5\times 10^4$ km/s in the case when the separation between them 
is $10^{-3}-10^{-2}$ pc \citep[see Table 1 in][]{pop12}. Such, double 
relativistic Fe K$\alpha$ lines and periodic X-ray variability are expected to 
be detected from very massive ($M>10^8\ M_{\odot}$) and cosmologically 
nearby ($z_{cosm} < 1$) SMBHBs \citep{ses12}.

The Fe K$\alpha$ line is produced by fluorescent emission from a very compact 
region around a SMBH \citep{fabi89,fabi00}, and thus it represents powerful 
diagnostic tool for studying physics and structure of such regions \citep[see 
e.g.][]{jova08,jova09,jova12,pop12}, as well as the masses and spins of SMBHs 
\citep[see e.g.][]{jova11,rey14}. X-ray reflection spectroscopy, which 
is based on the studies of the observed broad Fe K$\alpha$ line profiles, is 
nowadays proven to be an especially powerful technique for the robust black hole 
spin measurements across the wide range of black hole masses, from the 
stellar-mass black holes in the X-ray binaries to the SMBHs in the Active 
Galactic Nuclei (AGN), as well as for the reverberation of the relativistically 
broadened iron line which is already detected in the observed X-ray 
spectra of some AGN (for more details see the review by \citet{rey14}, and 
references therein).

In the case of SMBHBs, the line emitting regions could have different 
structures, depending on the mass ratios of the components, separation between 
them and parameters of their accretion disks \citep{jova14}. In some cases the 
secondary SMBH could be even embedded in the accretion disk around the primary, 
causing an empty annular gap in it \citep{mcke13}, similarly to the empty gap 
in circumbinary disk of Mrk 231 \citep[see Fig. 1 in][]{yan15}. Namely, 
it is well known that a SMBHB  with small mass ratio ($q\ll1$) can exchange 
angular momentum with its disk, distorting its density and causing the secondary 
SMBH to migrate inward \citep[see e.g.][and references therein]{mcke13}. If a 
secondary has a very low mass, it is subjected to a very rapid Type I migration, 
and it cannot significantly affect the Fe K$\alpha$ line emitted form the disk, 
neither can create an empty cavity in it. However, if the secondary is 
sufficiently massive (i. e. if the mass ratio of the SMBHB exceeds the critical 
value of $q \geq 10^{-4}$), it will be subjected to a Type II migration, and it 
will open an empty annular gap in the disk, analogous to the gaps in 
protoplanetary disks, significantly affecting the emitted Fe K$\alpha$ line 
\citep{mcke13}.

Here we study the possibility to detect the SMBHB signatures in their observed 
Fe K$\alpha$ line profiles by nowadays and future X-ray detectors. For this 
purpose we studied two models of the SMBHBs: (1) model with clearly separated 
components, where the accretion disks around both primary and secondary SMBHs 
significantly contribute to their composite line emission, and (2) model in 
which the secondary SMBH is embedded in the accretion disk around primary, 
causing an empty gap in the disk.

This paper is organized as follows: the simulations of X-ray radiation from two 
accretion disks with different parameters, as well as the procedure how to 
obtain the simulated Fe K$\alpha$ line profiles for two mentioned models of 
SMBHBs are described in Section 2, the obtained results are presented in 
Section 3 and briefly discussed in Section 4, and finally, in Section 5 we 
point out our main conclusions.

\section{Models of accretion disks around two components of a SMBHB and 
their barycentric orbits}

\subsection{Accretion disk models}

In order to study the potential observational signatures of SMBHBs in the Fe 
K$\alpha$ line profiles emitted from the relativistic accretion disks around 
their SMBH components, we performed the numerical simulations of disk emission 
based on ray-tracing method in Kerr metric, taking into account only those 
photon trajectories reaching the observer's sky plane 
\citep{fant97,cade98,jova08,jova09,jova12}. We assumed that disk emissivity 
$\varepsilon \left( {r} \right)$ follows power law: 
$\varepsilon\left({r}\right)=\varepsilon_{0}\cdot{r^p}$, where 
$\varepsilon_{0}$ is an emissivity constant, and $p$ is an emissivity index 
\citep[see][for more details]{jova12}. In the case of the SMBHB model with 
empty gap, this power law disk emissivity was modified so that 
$\varepsilon\left({r}\right)=0$ over the annulus representing the gap (see 
below text for more details).

Due to several effects, photons emitted from a disk at energy $E_{em}$ (or 
wavelength $\lambda_{em}$) will be observed by an observer at infinity at 
energy $E_{obs}$ (or wavelength $\lambda_{obs}$), causing the energy shift $g$ 
or, equivalently, the usual redshift in wavelength $z$ \citep[for more details 
see e.g.][and references therein]{jova16}:
\begin{equation}
\label{eqn:shift}
g=\dfrac{E_{obs}}{E_{em}}=\dfrac{\lambda_{em}}{\lambda_{obs}}=\dfrac{1}{1+z}.
\end{equation}
By integrating the observed flux at each observed energy over the whole disk 
image, one can obtain the corresponding simulated line profile emitted from the 
previously calculated accretion disk image \citep{jova12}:
\begin{equation}
\label{eqn:line}
F_{obs} \left( {E_{obs}} \right) = {\displaystyle\int\limits_{image} 
{\varepsilon\left({r} \right)}} g^{4}\delta \left( {E_{obs} - gE_{0}} 
\right)d\Xi
\end{equation}

\begin{figure}[t!]
\centering
\includegraphics[width=0.71\textwidth]{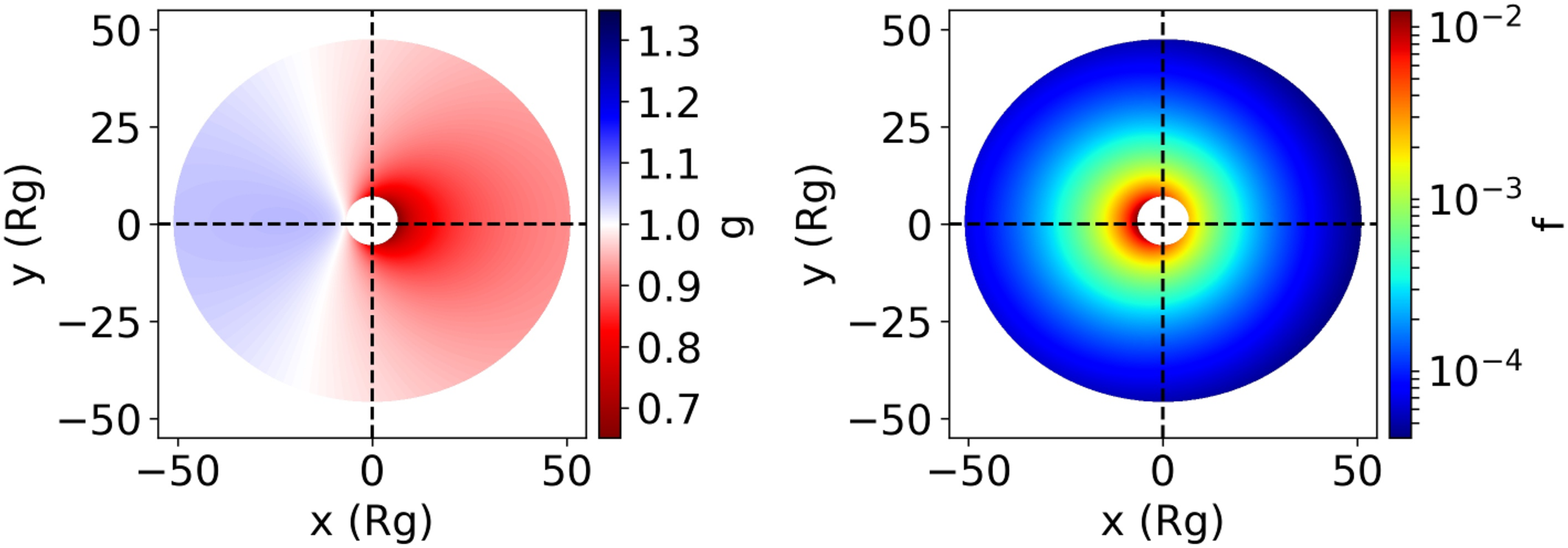}
\includegraphics[width=0.28\textwidth]{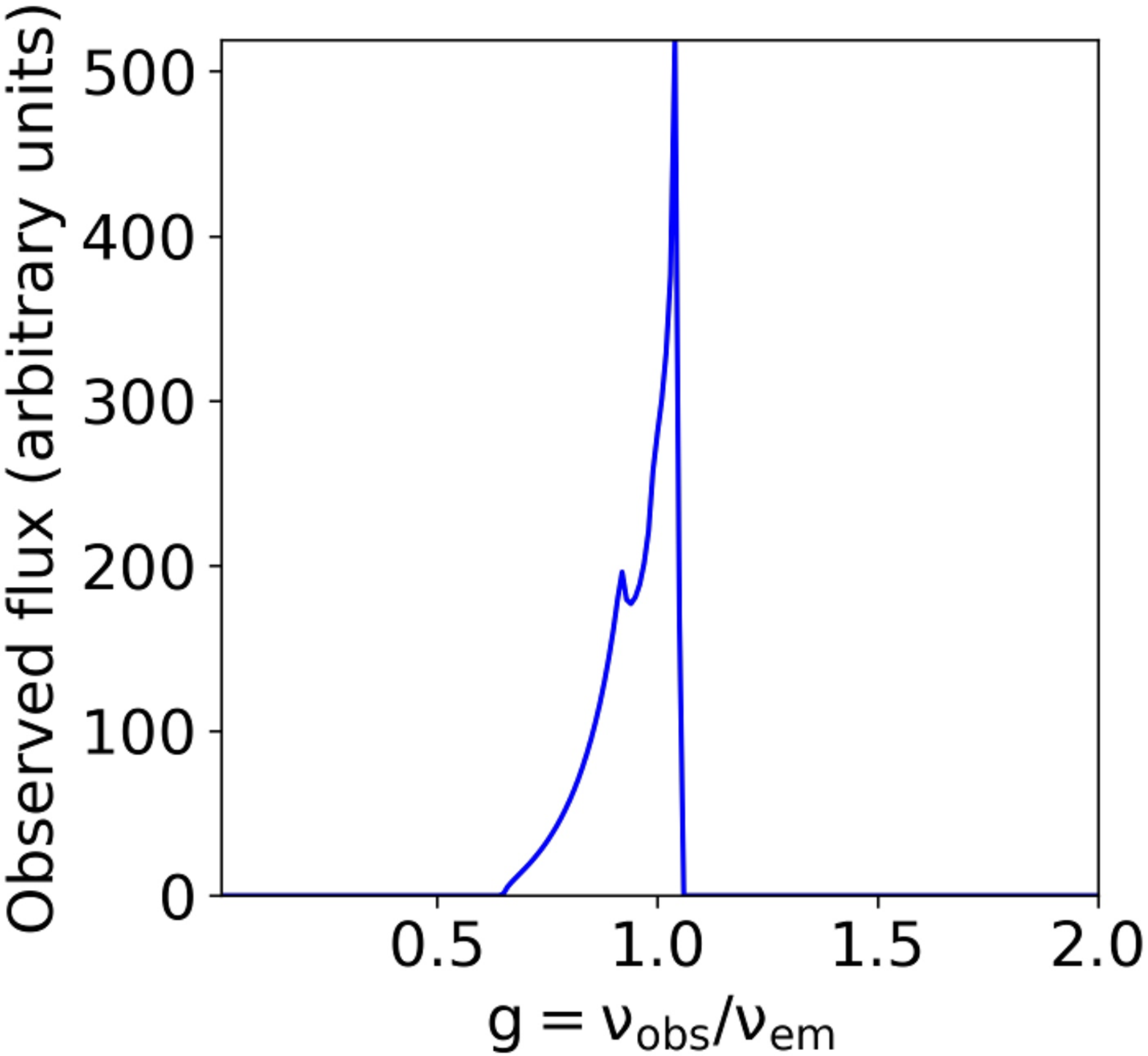}
\vspace{0.2cm} \\
\includegraphics[width=0.71\textwidth]{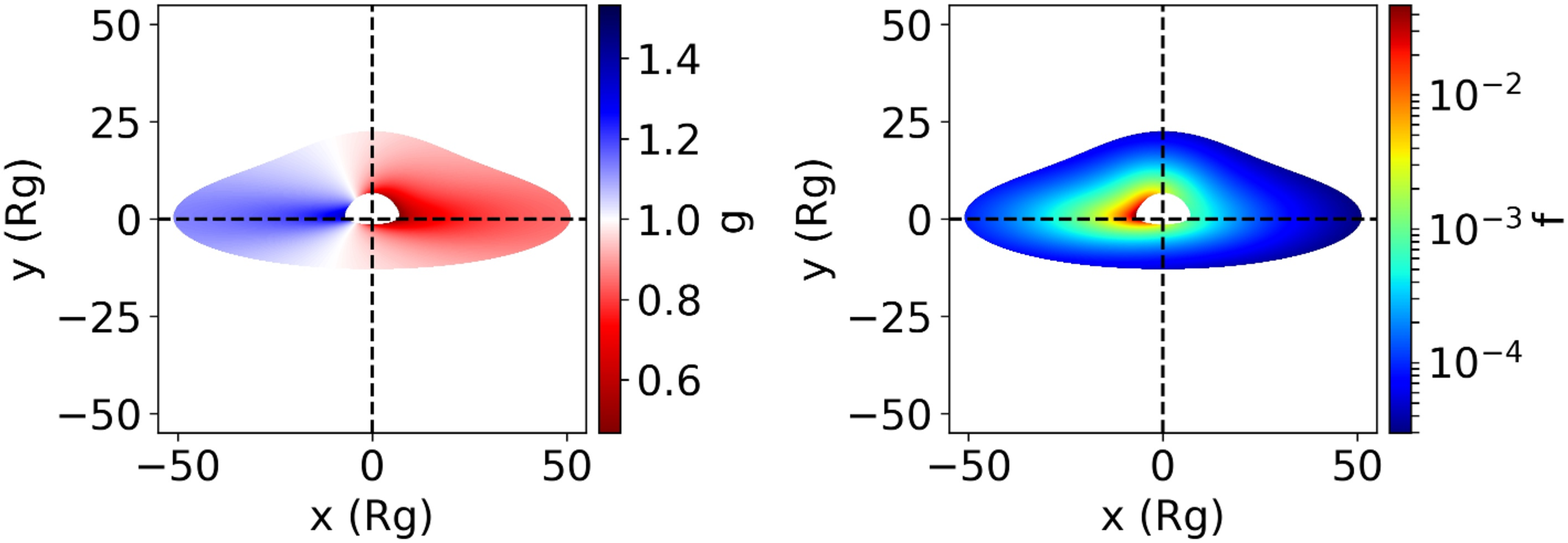}
\includegraphics[width=0.28\textwidth]{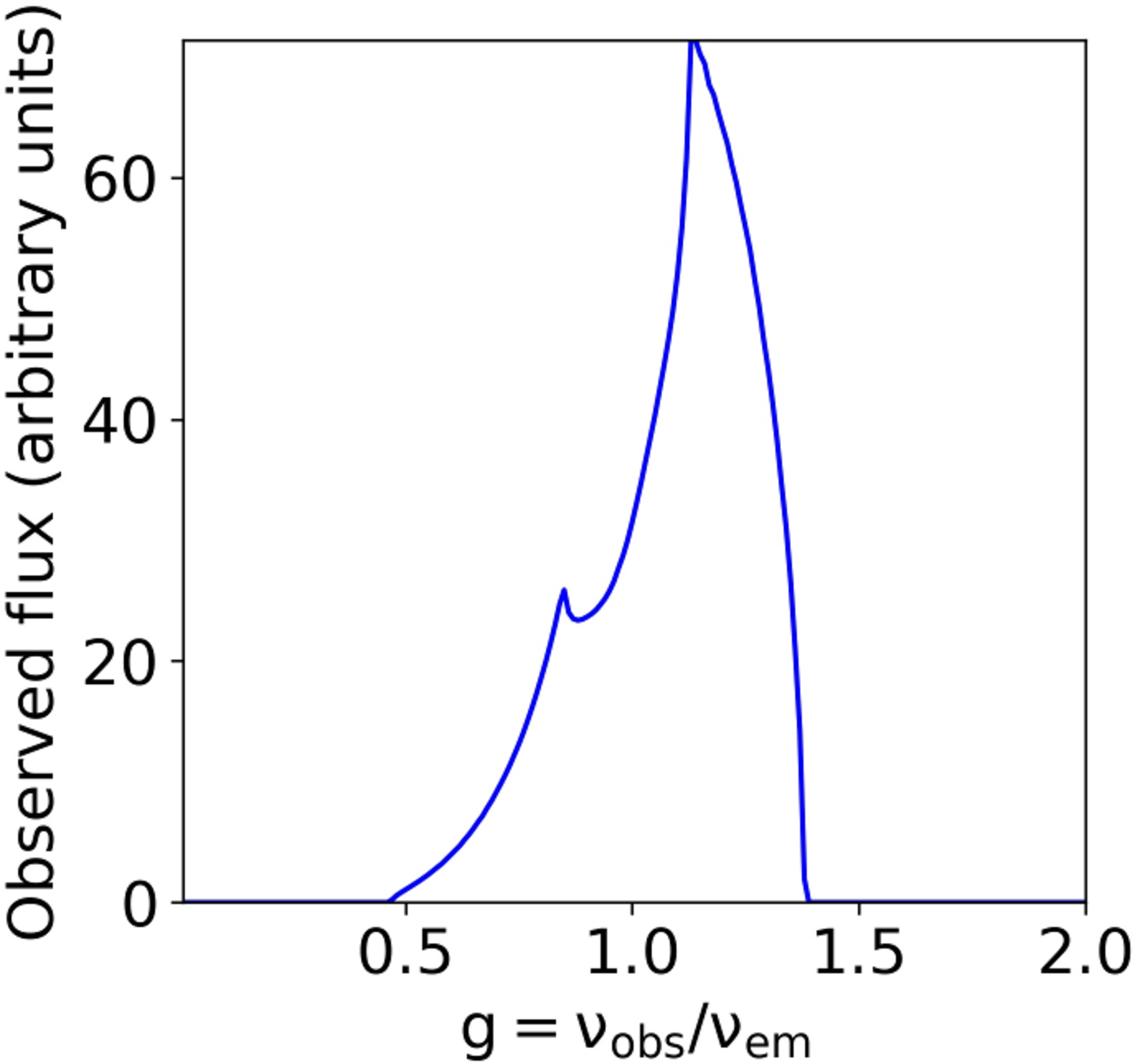}
\caption{Simulated images of the relativistic accretion disk in Kerr metric 
around a SMBH, colored according to energy shift $g$ (left) and observed flux 
(middle), as well as the corresponding simulated non-normalized line profile 
(right). Top panel corresponds to ``disk model A'' and bottom to ``disk model 
B'' (see \S 2 for the particular values of disk parameters).}
\label{fig01}
\end{figure}

Taking into account the results of some previous studies \citep[see e.g.][and 
references therein]{jova12}, we assumed the following parameters for modeling 
the accretion disks around the primary and secondary SMBHs: both SMBHs are 
assumed to be slowly rotating with the same small spin: $a_{BH} = 0.1$, inner 
radii 
of both disks are fixed to $R_{in} = R_{ms} = 5.67\, R_g$ (where $R_{ms}$ is 
the radius of the marginally stable orbit, and $R_g=GM/c^2$ - the gravitational 
radius of the SMBH with mass $M$), their outer radii are also assumed to be the 
same: $R_{out} = 50\, R_g$, as well as their emissivity indices: $p = -2.5$. We 
simulated the SMBHB signatures for two different disk inclinations: 
$\theta_{obs} = 
25^\circ$ (labeled as ``disk model A'' for the further reference) and 
$\theta_{obs} = 75^\circ$ (``disk model B''). The resulting simulated disk 
images, colored according to the energy shifts $g$ and the observed fluxes $F$, 
as well as the corresponding simulated non-normalized line profiles are 
presented in Fig. \ref{fig01} for both disk models. As it can be seen from this 
figure, disk inclination significantly affects the width and intensity of the 
resulting line profiles, which was the main reason why we assumed these 
particular two disk models.

Since, as already mentioned, the appearance of double relativistic Fe 
K$\alpha$ lines and periodic X-ray variability are expected to be detected from 
very massive and cosmologically nearby SMBHBs \citep{ses12}, we assumed 
a large mass of the primary SMBH: $M_1=5\times 10^8\ M_\odot$ and small 
angular diameter distance to the binary system of SMBHs: $D_A$ = 20 Mpc, 
corresponding to $z_{cosm}\approx 0.0045$ for $\Omega_M=0.315, 
\Omega_\Lambda=0.685$ and $H_0=67.4\ \mathrm{km\,s^{-1}\,Mpc}^{-1}$ 
\citep{pl18}. For mass ratio of the SMBHB we adopted the value $q=0.5$, 
corresponding to the mass of the secondary of $M_2=2.5\times10^8\ M_\odot$. 
Thus, the total mass of the SMBHB is almost an order of magnitude less than the 
mass of the central SMBH of M87, estimated from its shadow by Event Horizon 
Telescope Collaboration \citep{eht19}, and it is within the mass limit of 
$1.6\times 10^9\ M_\odot$ for the SMBHBs out to the distance of the Virgo 
Cluster ($\approx 16$ Mpc), obtained from 11-year data set for low frequency 
gravitational waves by NANOGrav PTA collaboration \citep{agg19}. This resulted 
with simulated accretion disk around the primary with outer radii of 
$R_{out}=50\ R_g\approx 250\ \mathrm{AU}$ which on the observer's sky 
corresponds to $\approx 12.5\ \mu\mathrm{as}$, giving a total apparent size of  
the disk's image of $\sim 25\ \mu\mathrm{as}\times 25\ \mu\mathrm{as}$.

Regarding the assumed size of the accretion disks, we took into account 
the results of some observational studies of AGN with single SMBHs in their 
centers, which suggest that the broad Fe K$\alpha$ line is emitted from the 
innermost regions of their accretion disks, extending between $R_{ms}$ and a 
few dozens of $R_g$ \cite[see e.g.][for an example in the case of radio galaxy 
4C 74.26]{bal05}. This is also in accordance with so called ''standard model of 
accretion disk'' \citep{sha73}, according to which the spectrum of radiation 
emitted from the disk depends on the distance to the central SMBH, so that the 
innermost part of the disk between $R_{ms}$ and several tens of $R_g$ emits 
X-rays, its central part between $\sim 100\ R_g$ and $\sim 1000\ R_g$ emits UV 
radiation, while its outer part located thousands $R_g$ from the SMBH, emits 
the optical radiation. Taking this into account, and since we are here 
investigating only the Fe K$\alpha$ line emission from a SMBHB, we assumed that 
the outer radii of its both disks are equal to the outer radii of their Fe 
K$\alpha$ line emitting regions, for which we adopted the following value: 
$R_{out}=50\ R_g$.

On the other hand, the outer parts of an accretion disk in a close 
SMBHB (i.e. when semimajor axis $a$ is small), could be subjected to the tidal 
disruption by the gravitational field of the second component. Therefore,
the maximum size of the accretion disks in such a SMBHB is determined by the 
gravitational interaction between its components, and it is a function of their 
mass ratio $q$ \citep[for more details see e.g.][]{pacz77}. An accretion disk 
with outer radius $R_{out}$ would be tidally disrupted if the semimajor axis 
$a$ of the SMBHB is less than the following limit for for tidal disruption 
\citep{saf19}:
\begin{equation}
\label{eqn:tidal}
a_t=R_{out}\left(1+q^{1/3}\right).
\end{equation}
In this study, the outer radii of the disks are $R_{out}\approx 250\ 
\mathrm{AU}$, which corresponds to the following limit for tidal disruption: 
$a_t\approx 450\ \mathrm{AU}$. As it can be seen from Table \ref{tab01}, the 
semimajor axes of both studied orbits significantly exceed this limit ($a\gg 
a_t$), and thus the accretion disks around both components are not subjected to 
the tidal disruption during their orbital motion.

\subsection{Keplerian barycentric orbits of SMBHBs}

The first of two studied SMBHB models assumes that its clearly separated primary 
and secondary components are moving around their common center of mass along 
Keplerian orbits which, due to their radial velocities, causes Doppler shifts in 
the Fe K$\alpha$ lines emitted from their accretion disks.

To model Keplerian barycentric orbits of a binary sistem of SMBHs, we 
apply the same procedure which is commonly used for the binary stars 
\citep[for more details see e.g.][]{hild01}. As a first step, we can 
adopt some masses of the primary and secondary components $M_1$ and $M_2$, or 
alternatively, just mass of the primary and mass ratio between the secondary 
and primary: $q=\dfrac{M_2}{M_1}$. We also need to assume some separation 
between the components $a$ (i.e. semimajor axis of their relative 
orbit)\footnote{Sub-parsec SMBHBs are of special interest for this 
investigation, since several observational studies indicated the existence of 
such SMBHBs in the cores of some AGN \citep[see e.g.][]{bon12}.}. The orbital 
period of the binary can be then obtained from the third Kepler's law:
\begin{equation}
\label{eqn:period}
P^2 = \dfrac{4{\pi ^2}{a^3}}{G\left(1 + q\right)M_1}.
\end{equation}
If we denote time with $t$ and time of the pericenter passage with $\tau$, then 
the next step is to calculate the mean anomaly $M$ (and also orbital phase 
$\Phi$): \begin{equation}
\label{eqn:manom}
M=\dfrac{2\pi}{P}\left(t-\tau\right)=2\pi\Phi .
\end{equation}
Assuming some orbital eccentricity $e$, the corresponding eccentric anomaly $E$ 
can be obtained by solving the Kepler's Equation:
\begin{equation}
\label{eqn:kepler}
M=E-e\sin{E},
\end{equation}
and the true anomaly $\theta$ can be calculated from $E$ according to:
\begin{equation}
\label{eqn:tanom}
\theta=2\arctan{\left(\sqrt{\dfrac{1+e}{1-e}}\tan{\dfrac{E}{2}}\right)}.
\end{equation}
Finally, the true barycentric orbits (i.e. those in the orbital plane) of the 
primary and secondary SMBHs are then represented by polar equations of the 
ellipse $r_{1,2}\left(\theta\right)$:
\begin{equation}
\label{eqn:torb}
r_{1,2}\left(\theta\right)=\dfrac{a_{1,2}\left(1-e^2\right)}{1+e\cos{\theta}},
\end{equation}
where $a_1=\dfrac{q\,a}{1+q}$ and $a_2=\dfrac{a}{1+q}$ are their semimajor 
axes\footnote{One should also take into account that the orientations of two 
barycentric orbits differ by 180$^\circ$ within the orbital plane.}.

The corresponding apparent orbits (in rectangular coordinates) can be 
calculated by projecting the true orbits on the observer’s sky plane using the 
remaining three Keplerian orbital elements (orbital inclination $i$, longitude 
of the ascending node $\Omega$ and longitude (or argument) of pericenter 
$\omega$):
\begin{equation}
\label{eqn:aorb}
\begin{array}{rl}
x_{1,2} = & r_{1,2}\cos{\theta}\left[\cos{\Omega}\cos{\omega}-\sin{\Omega}\sin{
\omega}\cos{i}\right] \\
+ & r_{1,2}\sin{\theta}\left[-\cos{\Omega}\sin{\omega}-\sin{\Omega}\cos{\omega}
\cos {i}\right] \\
& \\
y_{1,2} = & r_{1,2}\cos{\theta}\left[\sin{\Omega}\cos{\omega}+\cos{\Omega}\sin{
\omega}\cos{i}\right] \\
+ & r_{1,2}\sin{\theta}\left[-\sin{\Omega}\sin{\omega}+\cos{\Omega}\cos{\omega}
\cos {i}\right].
\end{array}
\end{equation}

Radial velocities of the components are given by the following expression 
\citep{hild01}: 
\begin{equation}
\label{eqn:vrad}
V_{1,2}^{rad}\left(\theta\right) = {K_{1,2}}\left[ {\cos \left( {\theta  + 
\omega } \right) + e \cdot \cos \omega } \right] + \gamma,
\end{equation}
where $K_{1,2}$ are the semiaplitudes of the velocity curves:
\begin{equation}
\label{eqn:samp}
K_{1,2} = \dfrac{{2\pi {a_{1,2}}\sin i}}{{P\sqrt {1 - {e^2}} }},
\end{equation}
and $\gamma$ is systemic velocity (which is assumed to be 0 km/s in our 
simulations). Assuming that $V_{1,2}^{rad} \ll c$, Doppler shifts in wavelength 
($z_{1,2}$) and energy ($g_{1,2}$) due to radial velocities of the components 
are given by:
\begin{equation}
\label{eqn:dshift}
z_{1,2}\approx 
\dfrac{V_{1,2}^{rad}}{c},\quad g_{1,2}=\dfrac{1}{1+z_{1,2}} .
\end{equation}
The total redshift factor $g_{tot}$, representing the net effect of both 
relativistic effects and radial velocities of the components, can be then 
obtained from Eqs. (\ref{eqn:shift}) and (\ref{eqn:dshift}):
\begin{equation}
\label{eqn:gtot}
g_{tot}=\dfrac{1}{1+z+z_{12}}=\dfrac{1}{\dfrac{1}{g}+\dfrac{1}{g_{1,2}}-1}.
\end{equation}
We studied the influence of Doppler shifts on the observed disk emission by 
calculating the simulated line profiles according to the expression 
(\ref{eqn:line}), but using the total redshift factor $g_{tot}$ instead of 
$g$.

\section{Results and discussion}

To study the observational signatures of the first SMBHB model, we 
performed simulations of the X-ray radiation from such a SMBHB with mass ratio 
$q=0.5$ for two different Keplerian orbits of its components, denoted as ``orbit 
1'' and ``orbit 2'', and determined by the orbital elements which are summarized 
in Table \ref{tab01}. As it can be seen from this table, the obtained 
orbital periods are on the order of a few years, which means that we simulated 
electromagnetic signatures in the Fe K$\alpha$ line of the nearby SMBHBs with 
semimajor axes of $\sim 10^{-3} - 10^{-2}\ \mathrm{pc}$, periods of a few years 
and masses above $\sim 10^8\ M_\odot$, and such SMBHBs are targets of present 
PTAs like NANOGrav, which search for their gravitational wave signatures 
\citep[see e.g.][]{agg19}.

\begin{table}[t!]
\centering
\caption{Adopted orbital elements for two Keplerian orbits of a SMBHB with 
mass ratio $q=0.5$.}
\begin{tabular}{|c|c|c|c|c|c|c|c|}
\hline
\hline
& \multicolumn{2}{c|}{$a$} & Period & $e$ & $i$ & $\Omega$ & $\omega$ \\
\cline{2-3}
& (AU) & (pc) &  (yr) & & $(^\circ)$ & $(^\circ )$ & $(^\circ )$ \\
\hline
\hline
\textit{orbit 1} & $1\times10^{3}$  & 0.005 & 1.16 & 0.5  & 30 & 0 & 30 \\
\hline
\textit{orbit 2} & $2\times10^{3}$  & 0.01 & 3.27 & 0.25 & 60 & 0 & 90 \\
\hline
\hline
\end{tabular}
\label{tab01}
\end{table}

\begin{figure}[t!]
\centering
\includegraphics[width=0.49\textwidth]{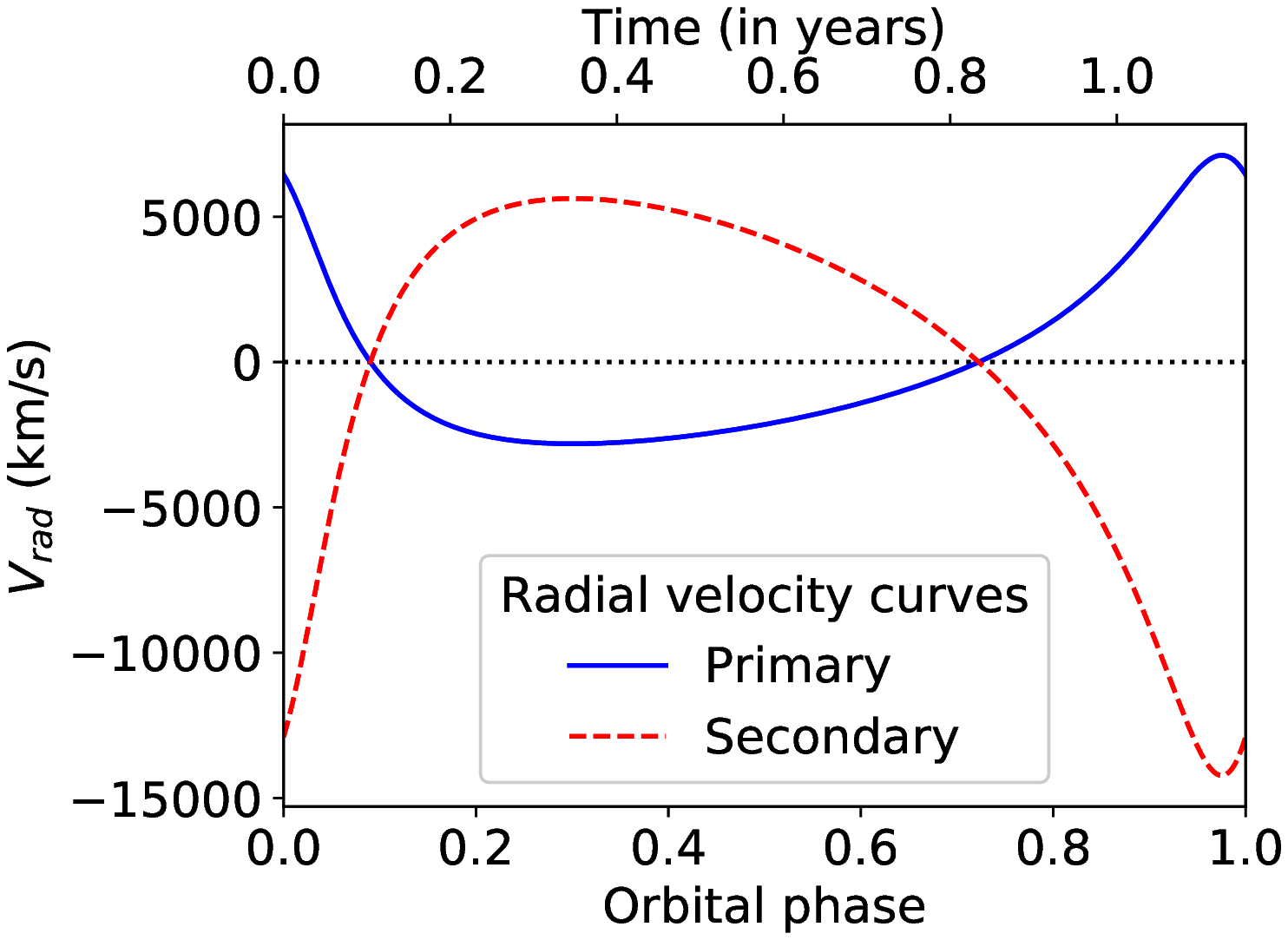}
\hfill
\includegraphics[width=0.49\textwidth]{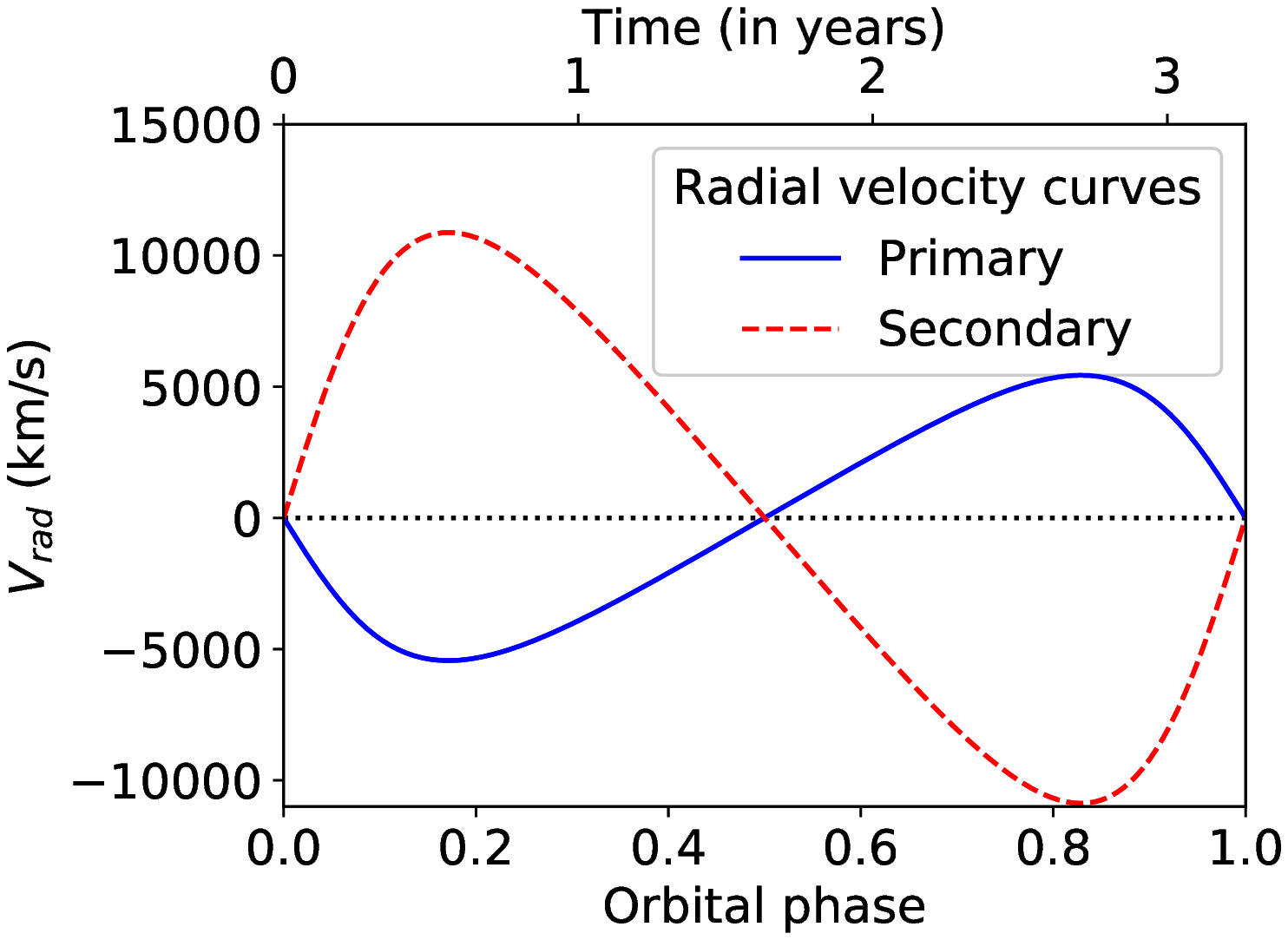}
\caption{Radial velocities of the components in a SMBHB with mass ratio $q$ = 
0.5, for orbit 1 (\textit{left}) and for orbit 2 (\textit{right}).}
\label{fig02}
\end{figure}

The obtained radial velocities of the components in the case of both orbits 
from Table \ref{tab01} are presented in Fig. \ref{fig02}. It can be seen from 
this figure that the radial velocity of the secondary SMBH can go far beyond 
10,000 km/s, which is sufficient to induce significant Doppler shift in the 
X-ray radiation from its accretion disk.

\begin{figure}[ht!]
\centering
\includegraphics[width=0.98\textwidth]{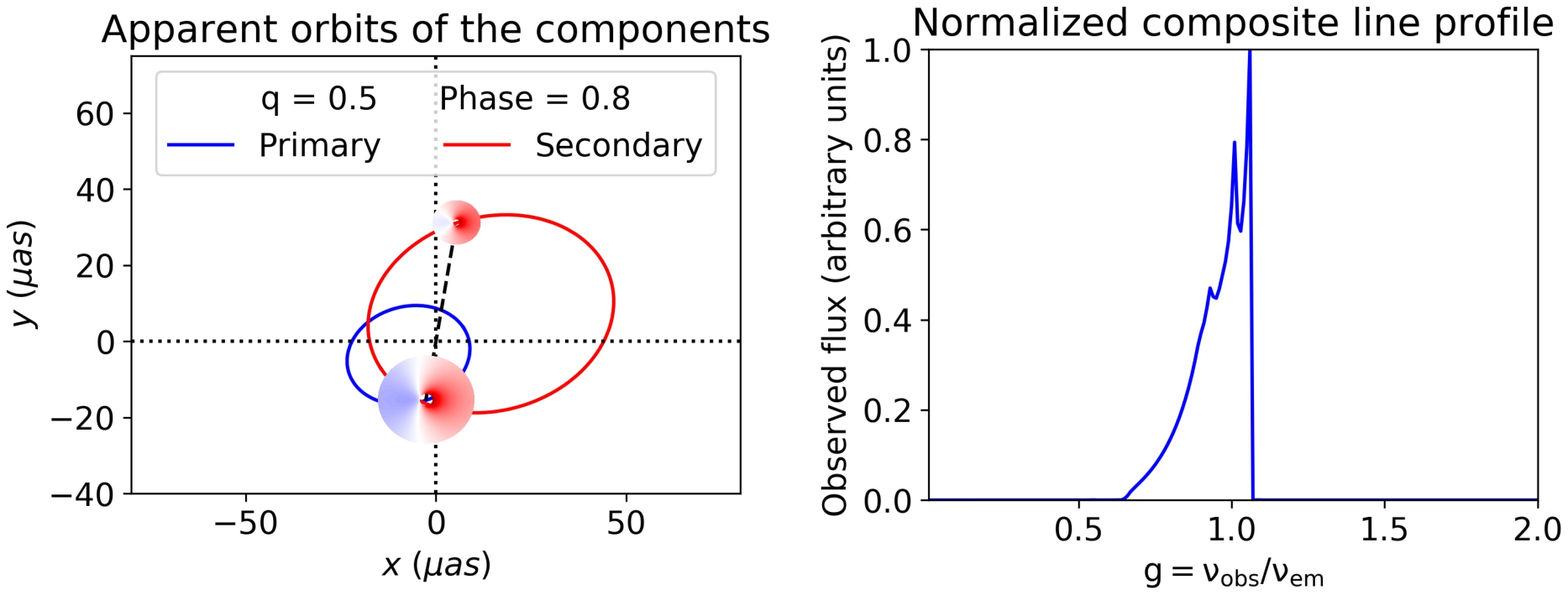}
\vspace{0.5cm} \\
\includegraphics[width=0.98\textwidth]{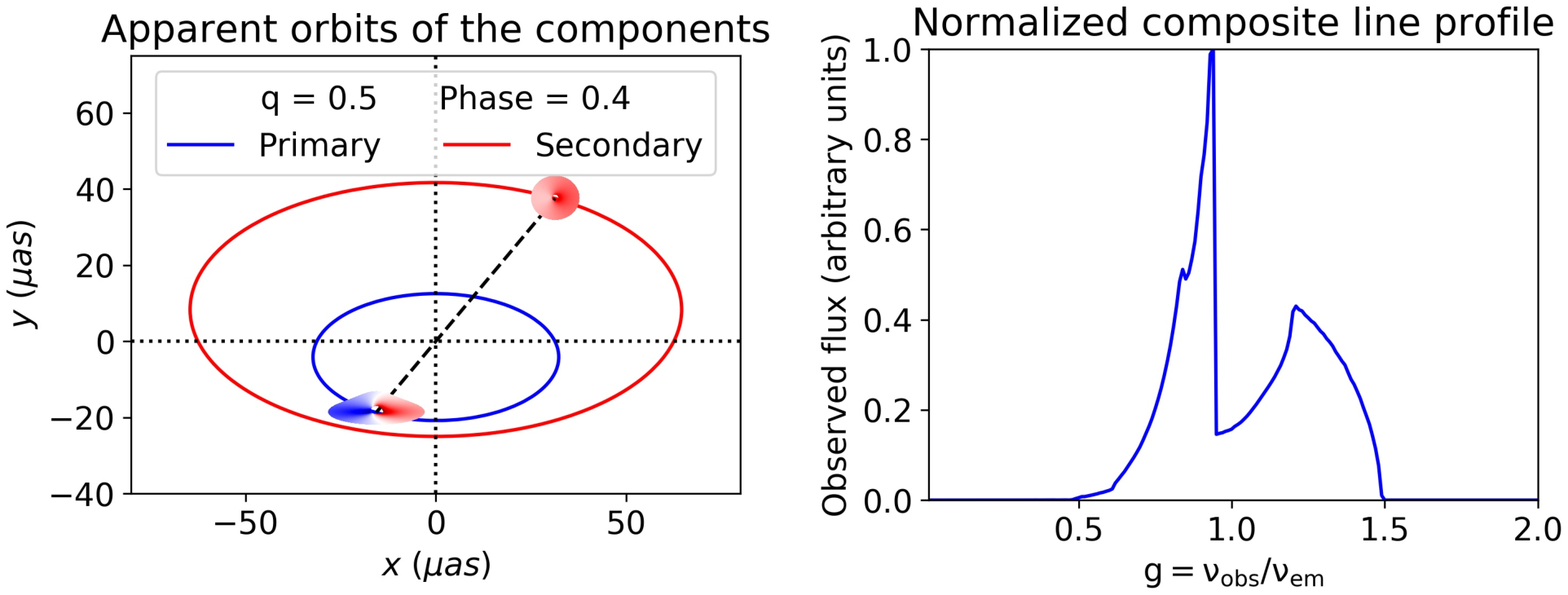}
\caption{\textit{Left}: Simulated images of the accretion disks around the 
primary and secondary components of a SMBHB with $q$ = 0.5 for ``disk A'' 
around both primary and secondary during orbital phase 0.8 along ``orbit 1''
(\textit{top panel}), as well as for ``disk B'' around primary and ``disk A'' 
around secondary during orbital phase 0.4 along ``orbit 2''  (\textit{bottom 
panel}). \textit{Right}: The corresponding simulated composite Fe K$\alpha$ 
line profiles.}
\label{fig03}
\end{figure}

As a next step, we used the obtained radial velocities to simulate the X-ray 
radiation from this SMBHB during different orbital phases along each of two 
orbits, assuming different models of accretion disks around its components. 
Two examples of the obtained results are presented in Fig. \ref{fig03} and they 
correspond to the orbital phase $\Phi=0.8$ along ``orbit 1'' in the case when 
``disk A'' is around both components (top panel), as well as to the orbital 
phase $\Phi=0.4$ along ``orbit 2'' in the case when ``disk B'' is around the 
primary and ``disk A'' around the secondary component (bottom panel). The disk 
images, colored according to the total redshift factor $g_{tot}$, are presented 
in the left parts of each panel of Fig. \ref{fig03}, while their right parts 
show the corresponding simulated composite Fe K$\alpha$ line profiles. As it 
can be seen from this figure, this model of SMBHB induces the appearance of the 
ripples in the cores of the simulated composite Fe K$\alpha$ line profiles. The 
amplitudes of these ripple effects strongly depend on the assumed parameters of 
the accretion disks around both primary and secondary SMBHs, the orbital 
elements of their Keplerian orbits, as well as on the particular orbital phase 
(i.e. time) at which, here a simulation and in reality an observation, is made.
Moreover, as demonstrated in \citet{jova14}, the mass ratio $q$ also 
has a strong influence on the amplitudes of these ripples, since the ratio 
between the fractions of the composite Fe K$\alpha$ line flux, contributed by 
the accretion disks around the secondary and the primary, roughly corresponds 
to $q$. Since a highly inclined disk produces a wide line profile, as 
demonstrated in Fig. \ref{fig01}, the ripple in the bottom panel of Fig. 
\ref{fig03} is much wider and deeper than the one in the top panel because it 
is created in the component of the Fe K$\alpha$ line which is emitted from the 
highly inclined ``disk B'' around the primary SMBH. 

\begin{figure}[ht!]
\centering
\includegraphics[height=0.45\textheight]{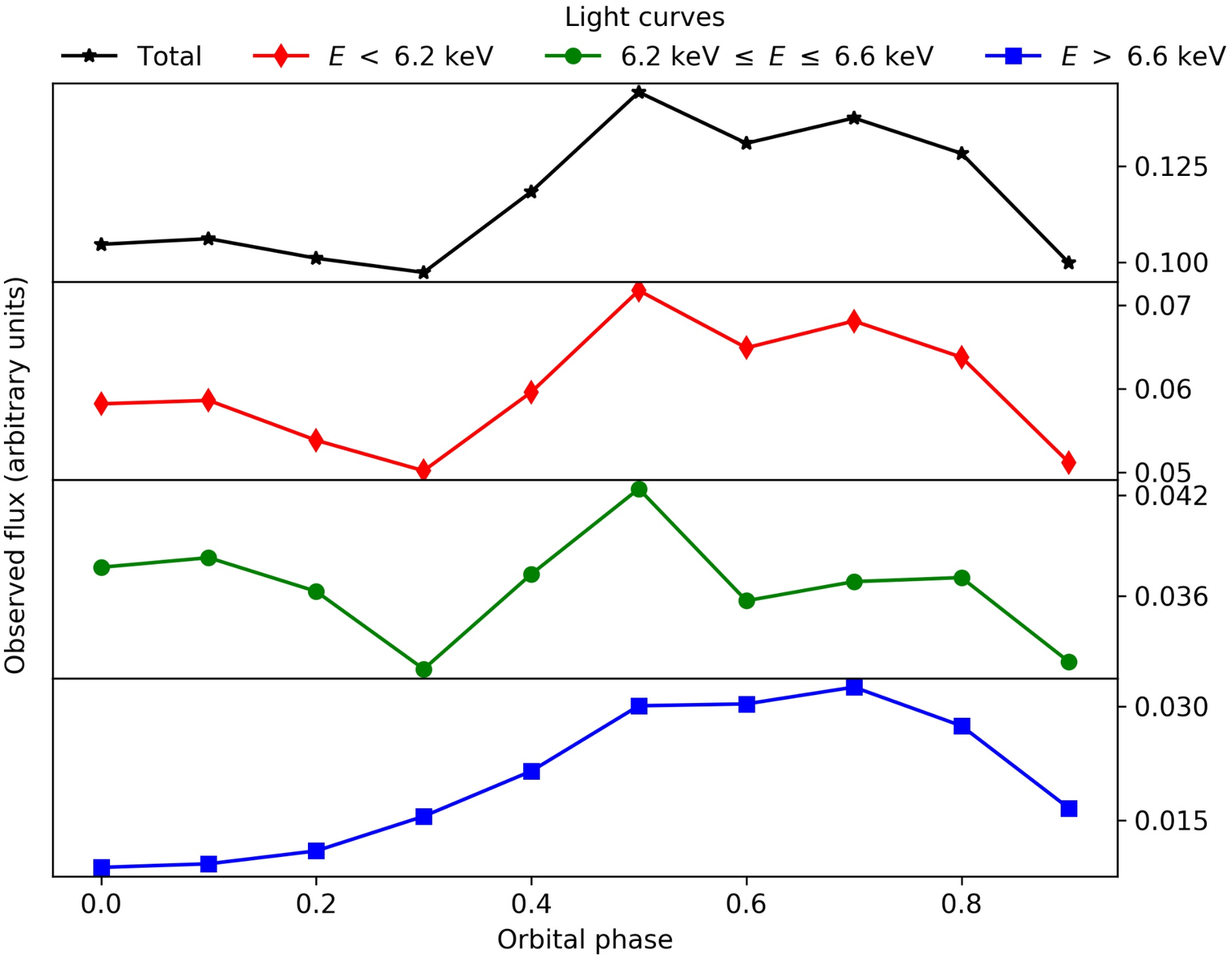}
\vspace{0.5cm} \\
\includegraphics[height=0.45\textheight]{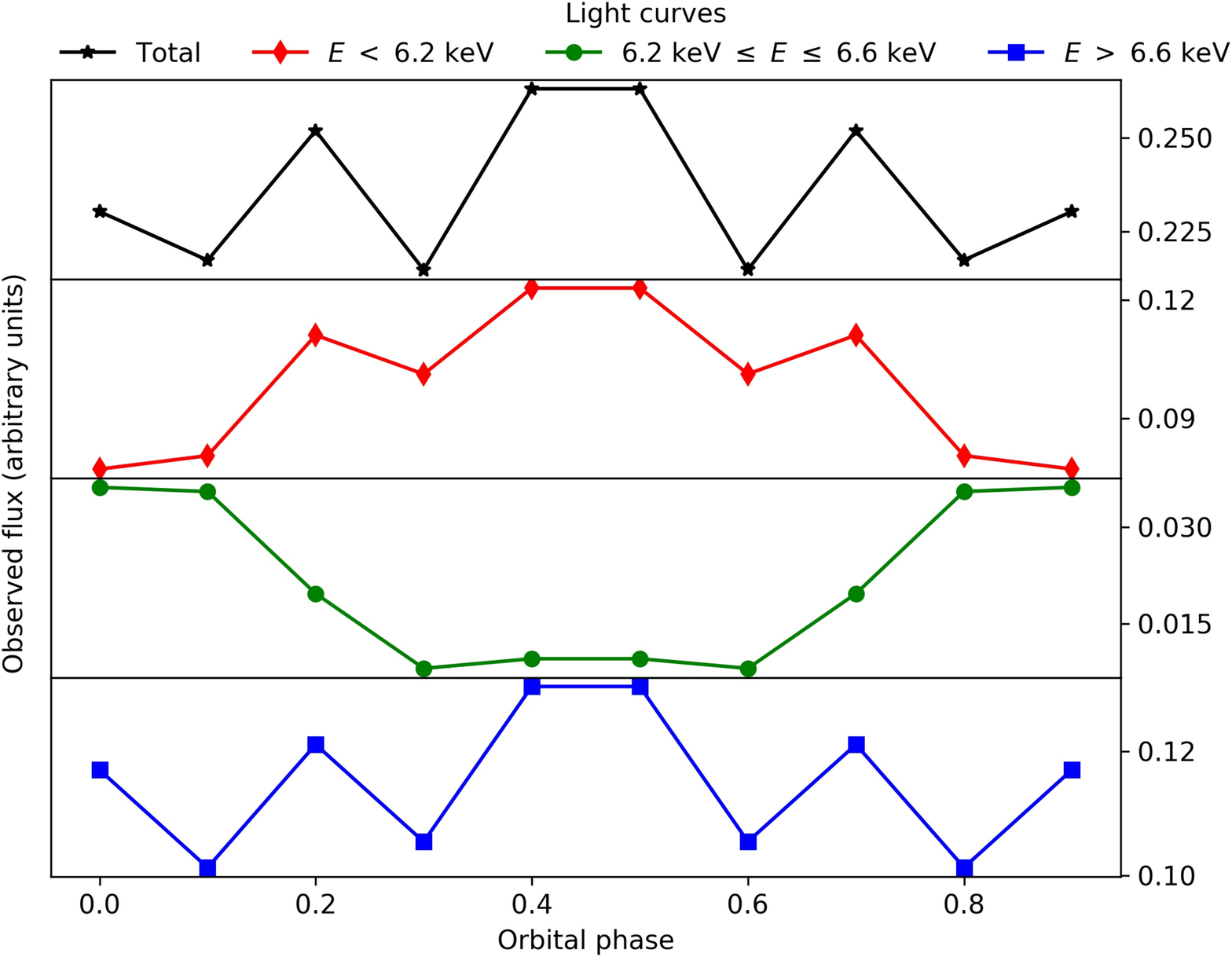}
\caption{Flux variability (from top to bottom) of the total Fe K$\alpha$ line 
profile (black line), 
its red part (red line), core (green line) and blue part (blue line) during 10 
different orbital phases and for the same disk parameters and orbital elements 
as in the corresponding panels of Fig. \ref{fig03}.}
\label{fig04}
\end{figure}

We also investigated the flux variability in different parts of the Fe 
K$\alpha$ line for different phases during one orbital period. The flux 
variability of the total line profile, its red part ($E < 6.2$ keV), core (6.2 
keV $\le E \le 6.6$ keV) and blue part ($E > 6.6$ keV) are presented in Fig. 
\ref{fig04} by black, red, green and blue lines, respectively. Two panels in 
Fig. \ref{fig04} refer to the same disk parameters and orbital elements as 
the corresponding panels in Fig. \ref{fig03}, but for 10 different orbital 
phases (including those from Fig. \ref{fig03}). As it can be seen from Fig. 
\ref{fig04}, this model of SMBHBs induces the highest flux variability in 
the core of the Fe K$\alpha$ line. Taking into account that the presented 
results refer to only one orbital period, it can be easily deduced that the 
same variability pattern would also repeat during all successive orbital 
periods. Therefore, in the case of such a SMBHB, a periodic variability of the 
X-ray radiation in the Fe K$\alpha$ line should be expected, which is in good 
agreement with the similar predictions of \citet{ses12}. 

\clearpage

\clearpage

\begin{figure}[ht!]
\centering
\includegraphics[width=0.90\textwidth]{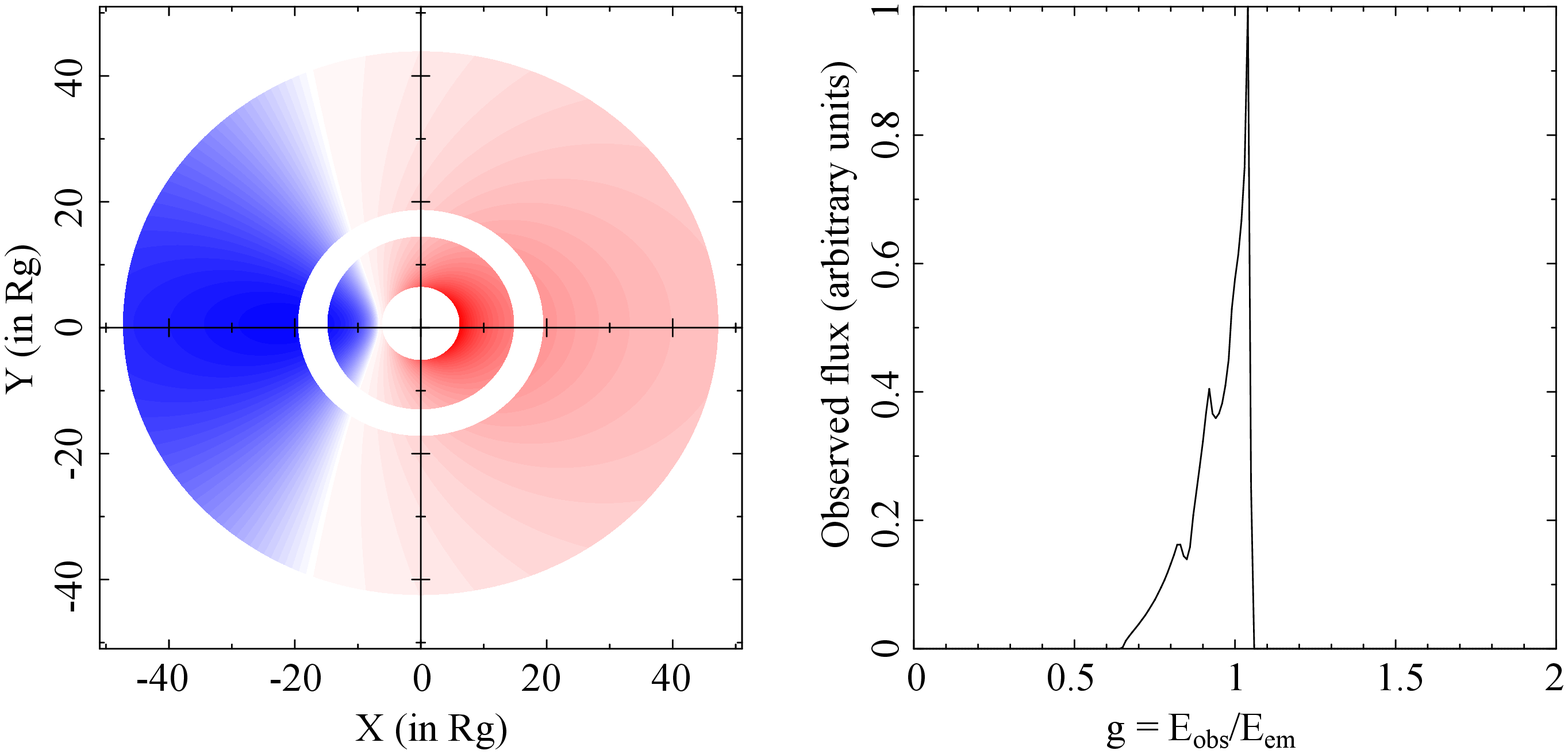}
\vspace{0.5cm} \\
\includegraphics[width=0.90\textwidth]{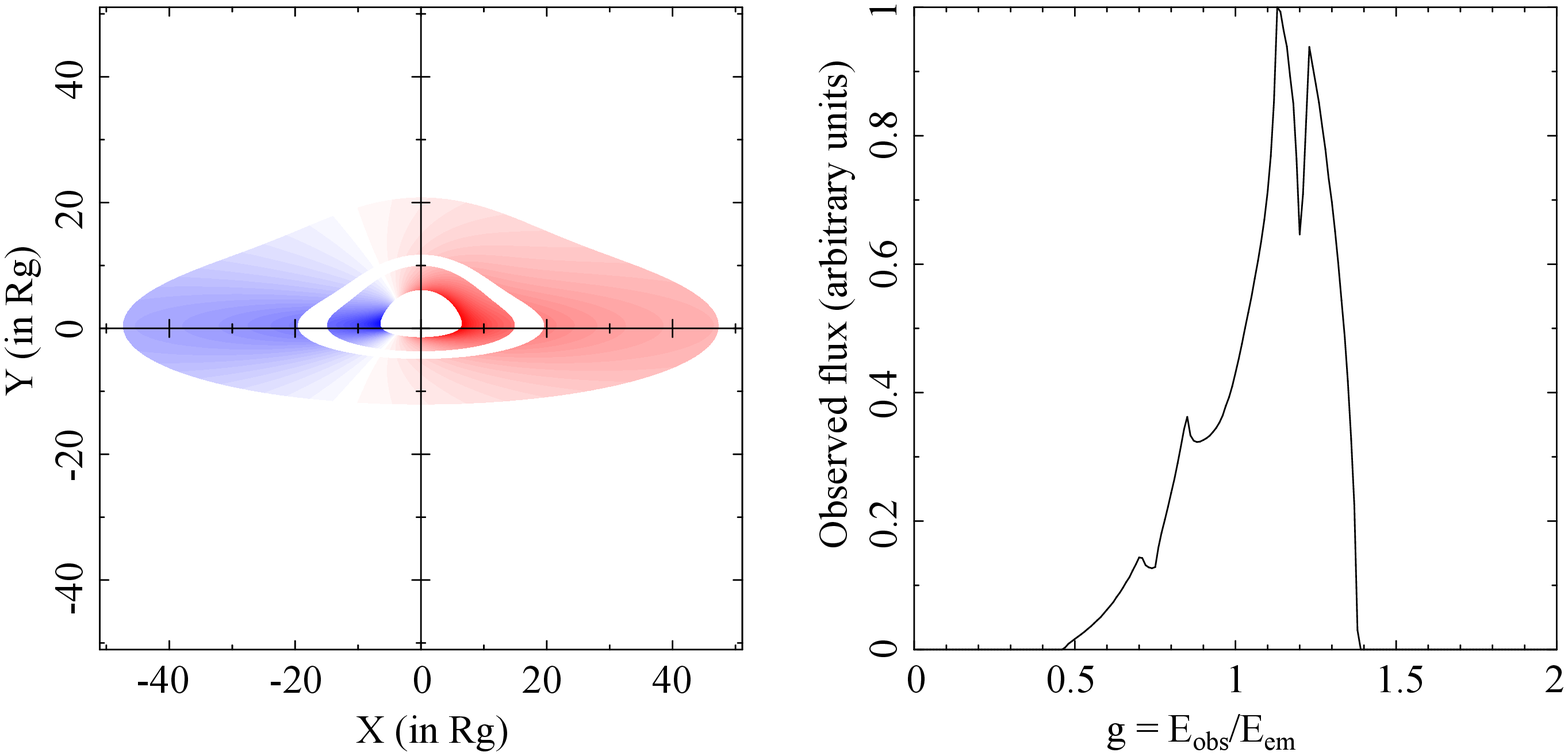}
\caption{Illustrations of two relativistic accretion disks with an empty gap 
extending from $15 R_g$ to $20 R_g$ (\textit{left}), and the corresponding 
simulated Fe K$\alpha$ line profiles (\textit{right}). Top panel corresponds to 
the first, and bottom panel to the second model of accretion disks, as 
described 
in $\S$2.}
\label{fig05}
\end{figure}

We studied the observational signatures of the second SMBHB model using the 
similar simulations of the X-ray radiation from the accretion disk around its 
primary SMBH, in which its secondary SMBH cleared an empty gap. We performed 
these simulations for both disk models from \S2 and for different positions and 
widths of the empty gap, i.e. for different values of its inner and outer 
radii. Two examples of the obtained results in the case of ``disk A'' and 
``disk B'', each with an empty gap extending between $15\ R_g$ and $20\ R_g$, 
are shown in the left panels of Fig. \ref{fig05}, while the corresponding 
simulated Fe K$\alpha$ line profiles are shown in the right panels of the same 
figure. Taking into account that the gap is formed in the accretion 
disk around the primary, and for the sake of simplicity, we expressed its 
radii in the units defined by gravitational radius $R_g=R_{g1}$ (and mass) of 
the primary SMBH, in which the both gaps from the left panels of Fig. 
\ref{fig05} have the widths of $5\ R_{g1}$. On the other hand, the gap width 
can be also expressed in different units of gravitational radius $R_{g2}$ for 
the secondary SMBH (or equivalently, in terms of its mass), assuming that it is 
twice as large as the outer radius of the accretion disk around the secondary, 
for which we adopted the value of $50\ R_{g2}$, as previously mentioned. Under 
these assumptions: $5\ R_{g1}=100\ R_{g2}$, resulting with the following value 
for the mass ratio of the SMBHB: $q=\dfrac{R_{g2}}{R_{g1}}=0.05$, as well as 
with the mass of the secondary SMBH of $M_2=2.5\times 10^7\ M_\odot$. 
Therefore, this SMBHB has an order of magnitude lower mass ratio than the 
previously described model, but which is still significantly larger than the 
limit of $q=10^{-4}$ below which the secondary SMBH is not able to open a gap 
in the disk around the primary \citep[see e.g.][for more details]{mcke13}. 
Another important difference between these two models is an assumption that in 
the latter case, there is no Fe K$\alpha$ line emission from the gap, i. e. 
that the ''minidisk'' around the secondary SMBH (which is embedded in the gap) 
gives a negligible contribution to the total line emission of the binary. Such 
scenario is preferable, although one cannot completely exclude a possibility 
that some gas can enter the gap and feed a ''minidisk'' around the secondary 
which, as a consequence, could create some additional Fe K$\alpha$ line 
emission \citep[see][and references therein]{mcke13}.

It can be seen from Fig. \ref{fig05} that in such a case an empty 
annular gap in the accretion disk around primary could have significant 
influence on the resulting line shape, inducing the appearance of the similar 
ripples as in the case of the first SMBHB model, which is in good agreement with 
the results of some previous studies \citep[see e.g.][]{mcke13}. However, these 
ripple effects in the second SMBHB model, besides having the lower amplitudes, 
strongly depend only on the positions and widths of empty gap (or in other 
words, on its radii), but they do not noticeably vary with time (or orbital 
phase), since we assumed that the inner and outer radii of the gap are also time 
independent. In reality, on the other hand, the secondary SMBH would slowly 
spiral down towards the primary, so the radii and width of empty gap would also 
change with time, inducing a slow time variability of the resulting ripple 
effects.

\section{Discussion}

As it was shown in the previous section, both studied SMBHB models predict 
occurrences of the ripples in the cores of emitted Fe K$\alpha$ line profiles. 
Thus, an important related issue is the possibility to detect such effects in 
the observed spectra of SMBHB candidates using nowadays and future X-ray 
detectors, which depends on their spectral resolution and signal-to-noise ratio 
(S/N).

The presented simulated line profiles were calculated over 200 bins of width
$\Delta E = 0.064$ keV, which corresponds to the spectral resolution of $E/ 
\Delta E = 100$ at 6.4 keV. On the other hand, the spectral resolutions  
of some modern X-ray observatories are: $E/ \Delta E\sim 20-50$ in the case of 
XMM-Newton, $\sim 600$ at 6 keV for Suzaku, and $\sim 100-1000$ in 0.1--10 keV 
range for Chandra. Therefore, the performed simulations provide spectral 
resolution which is comparable to those of modern X-ray detectors, and 
according just to this property, the SMBHB signatures in the form of ripple 
effects in their observed Fe K$\alpha$ line profiles should be detectable even 
by nowadays X-ray detectors.

However, this is not achievable yet due to insufficient S/N of the modern 
instruments. Namely, in the ray-tracing simulations S/N depends on the number 
of photons emitted from a simulated disk \citep{milo18}, which in our case was 
5000 $\times$ 5000. Such a large number of photons in our simulations provides 
much higher S/N ratio than in the current observations by XMM-Newton and 
Chandra, which is the main cause for difficult detection of SMBHB signatures in 
the currently observed line profiles, in spite of similar spectral resolutions. 
However, the planned future X-ray observatories (like Advanced Telescope for 
High Energy Astrophysics -- ATHENA) will be equipped with high signal-to-noise 
and high spectral resolution instruments ($E/ \Delta E \sim 2800$ in 0.2--12 
keV range) and will enable the detection of the SMBHB signatures in the 
observed Fe K$\alpha$ line profiles.

\section{Conclusions}

We simulated the Fe K$\alpha$ line profiles emitted from the following two 
models of SMBHBs:
\begin{itemize}
\setlength{\itemsep}{0pt}
\item[(i)] a model in which both primary and secondary SMBHs are surrounded by 
an accretion disk and they are orbiting around their center of mass, and
\item[(ii)]  a model in which the secondary SMBH clears an empty gap (or cavity) 
in the disk around primary.
\end{itemize}
We can summarize the obtained results of these simulations as follows:
\begin{itemize}
\setlength{\itemsep}{0pt}
\item[(i)] Both models leave detectable ripples in the emitted Fe K$\alpha$ 
line profiles which may look like an absorption component in the line profile;
\item[(ii)] In the first model, such ripples in the composite line profiles are 
caused by Doppler shifts due to orbital motion, and depend on:
\begin{itemize}
\setlength{\itemsep}{0pt}
\item orbital phase of SMBHB (time) and cause the periodical variability of the 
line shapes,
\item mass ratio between the secondary and primary SMBHs,
\item parameters of the accretion disks (e.g. inclination) around both primary and secondary,
\item Keplerian orbital elements, which could potentially enable reconstruction 
of the observed radial velocity curves and their fitting with Keplerian orbits 
\citep[see e.g.][for an example]{bon12};
\end{itemize}
\item[(iii)] In the second model, these ripples do not significantly change in 
time, but instead:
\begin{itemize}
\setlength{\itemsep}{0pt}
\item they depend on the parameters of the disk around the primary 
(especially on the disk inclination),
\item their amplitudes strongly depend on the width and distance of the empty 
gap from the central SMBH, and hence they could be used for constraining the 
mass ratios and separations between the components in this type of SMBHBs;
\end{itemize}
\item[(iv)] Spectral resolutions and, especially S/N, of 
modern X-ray detectors are not sufficient to study in details such signatures 
of SMBHBs, however this will be possible with the next generation of X-ray 
observatories, such as ATHENA.
\end{itemize}

\acknowledgements
This work has been supported by Ministry of Education, Science and 
Technological Development of the Republic of Serbia, through the projects: 
176003 ''Gravitation and the Large Scale Structure of the Universe'' and 176001 
''Astrophysical Spectroscopy of Extragalactic Objects''. The authors 
would like to thank an anonymous referee for very useful and helpful comments 
and suggestions which significantly improved the presentation of the paper.

\end{document}